\documentstyle[12pt,epsfig]{article}

\title{
{\Large \bf Neutrino mixing schemes and
neutrinoless double beta decay\\}
\vspace*{0.6cm}
}

\author{H.V. Klapdor-Kleingrothaus$^{a\,}$,
and U. Sarkar$^{a\,b\,}$\\[0.5cm]
\normalsize $a$: {\it Max-Planck-Institut f\"ur Kernphysik, Postfach
  10 39 80,}\\[-0.15cm]{\it D-69029 Heidelberg, Germany}\\
\normalsize $b$: {\it Physics Department, Visva-Bharati
  University,}\\[-0.15cm]
{\it Santiniketan - 731 235, India}\\[2cm]
}
\date{}
\begin{document}
\maketitle
\thispagestyle{empty}
\vspace*{-2cm}

\begin{abstract}\

We study the possible structure of the neutrino mass matrix
taking into consideration the solar and atmospheric neutrinos and the
neutrinoless double beta decay. We emphasize
on mass matrices with vanishing elements. There are only a very
few possibilities remaining at present. We concentrate on three
generation scenarios and find that with three parameters there are few
possibilities with and without any vanishing elements. For
completeness we also present a five parameter four neutrino (with one
sterile neutrino) mass
matrix which can explain all these experiments and the LSND result.

\end{abstract}

\newpage
\baselineskip 18pt

In recent past there have been many new results in neutrino physics
\cite{atm}-\cite{ndbex}.
All these results are narrowing down the parameter space for the
neutrino masses and mixing. The solution to the atmospheric neutrino
anomaly \cite{atm} is the strongest to constrain the mass squared
difference and the mixing angle
between the $\mu$ and the $\tau$ neutrinos. This result is also
supported by the K2K result \cite{k2k} and a combined analysis
of both these experiment can now be used to determine the allowed
parameter range \cite{fl}. The maximal mixing angle
restricts the structure of the mass matrix very strongly.
Another very strong constrain comes from the CHOOZ result \cite{chooz}
from the non-observation of oscillation of the electron neutrino
into any other neutrinos. For the solar neutrinos
\cite{sol} there are a few possible solutions,
which are also narrowing down \cite{bah1}.
The small angle solution of the solar neutrinos is almost ruled
out. The situation with the sterile neutrinos is even
worse. It is not considered to be a favoured solution to the atmospheric
neutrinos and also for the solar neutrinos. However, for the
consistency of the LSND result \cite{lsnd}
with both the solar and the atmospheric
neutrinos we need a sterile neutrino. Thus, although the
present popular scheme is
to ignore the LSND result and work with a three generation
neutrino mass matrix, for completeness we shall also mention a four
generation scenario. The three mixing angles of a three generation
mass matrix is determined by the mixing angles required for the
atmospheric neutrinos, solar neutrinos and the reactor constraint from
CHOOZ.

Although the mass squared differences are required to be fairly small
compared to what is required for the neutrinos to contribute to the
dark matter of the universe \cite{gasilk},
the overall masses of the neutrinos could
be large enough to constitute the hot dark matter component of the
universe, which is required to explain the large scale structure of
the universe. The recent indication of the positive evidence for a
neutrinoless double beta decay \cite{ndbex} points exactly
to this interesting solution of the neutrino masses \cite{old,imp,impn}.
This result has now generated some interest in the field \cite{vis,new}.
If we include this result, the mass matrix
for the three generation scenario will have a very little freedom now.

Among other things, there is now a pattern in the mass matrix
\cite{alt,mix,vis,xx}.
Consider a three generation scenario in the flavour basis, when the
charged lepton mass matrix is diagonal $M^{\ell^-} = {\rm Diag}[m_e,
m_\mu, m_\tau]$. In this case the exact maximal
mixing for the atmospheric neutrinos implies that the four elements
$M^\nu_{22}, M^\nu_{33}, M^\nu_{23}$ and $M^\nu_{33}$ are
non-vanishing and constrained. It also implies that $M^\nu_{12} =
M^\nu_{13}$. The solar neutrino mixing angle then requires $M^\nu_{12}$
and $M^\nu_{13}$ to be non-vanishing. Finally $M^\nu_{11}$ gives the
contribution to the neutrinoless double beta decay. This discussion
shows that the neutrino mass matrix is completely determined now and
we have very little freedom left. Moreover all the elements are
required to be non-vanishing. However, since the atmospheric neutrino
mixing could be less than maximal and if all uncertainties in the
allowed parameters are considered this simple argument does not work
exactly.

Assuming the symmetric neutrino
mass matrix to be real there are six parameters. All these six parameters are
required to explain the atmospheric
neutrino anomaly, solar neutrino problem, the neutrinoless double
beta decay and satisfy the CHOOZ constraint. The explanation of the hot dark
matter is coupled with the neutrinoless double beta decay result, so
this is not considered as an independent constraint. The question we
would now like to ask is, can we have a neutrino mass scheme
satisfying all the present experimental constraints with less than six
parameters? What is the minimum number of parameters we require for a
neutrino mass matrix, which can satisfy all these constraints. In
particular, can we afford to have any elements of the neutrino mass
matrix to be vanishing. Exact zero elements in a mass matrix always
makes it convenient for model building, so we try to find out if there
exists any neutrino mass matrix with zero elements satisfying all
these constraints. As a sequel we present a four generation neutrino
mass matrix, which satisfies all these results including the LSND result
with only five (or even four) parameters.

The various
experimental inputs we consider are the following. The neutrinoless
double beta decay experiment constrain the $(11)$ element of the
neutrino mass matrix ($m_{ee}$) in the flavour basis (in which the
charged lepton mass matrix is diagonal). We define the mixing matrix
$U_{\alpha i}$ to be the one relating the physical states
$|\nu_\alpha>$ (in the flavour basis $\alpha = e,\mu,\tau$)
to the mass eigenstates $|\nu_i >$ (with masses $m_i$, $i=1,2,3$)
$$ |\nu_\alpha > ~ = ~ \sum_i ~ U_{\alpha i} ~ |\nu_i > .$$
Then the present evidence of the neutrinoless double beta decay gives
\begin{equation}
m_{ee} = \sum_i |U_{e i} |^2 m_i =
(0.05 - 0.86) ~ {\rm eV} ~ {\rm (at~ 97 \% ~ c.l.)}
\end{equation}
with a best fit $m_{ee} = 0.39$ eV.

The other constraints on the mass
eigenvalues are from the atmospheric and the solar neutrinos and the LSND:
\begin{eqnarray}
\Delta m^2_{atm} &=& \{(1.5 - 4.8), 2.7\} \times 10^{-3} ~ {\rm eV}^2 \nonumber
\\
\Delta m^2_{sol-LMA} &=& \{(2 - 30), 4.5\} \times 10^{-5} ~ {\rm eV}^2 \nonumber
\\
\Delta m^2_{sol-SMA} &=& \{(4 - 6), 4.7\} \times 10^{-6} ~ {\rm eV}^2 \nonumber
\\
\Delta m^2_{LSND} &>& 0.2 ~ {\rm eV}^2 . \nonumber
\end{eqnarray}
The last numbers are the best fit values.
For solar neutrinos the large mixing angle MSW solution (LMA) is the
preferred solution. However, we mention the small mixing angle MSW
solution (SMA) for completeness, which is almost ruled out
at the 2$\sigma$ level.
We shall not discuss the SMA solutions in detail. The corresponding
mixing angles are:
\begin{eqnarray}
\sin^2 2 \theta_{atm} &=& \{(0.87 - 1), 1\} \nonumber
\\
\sin^2 2 \theta_{sol-LMA} &=& \{(0.3 - 0.94), 0.82\} \nonumber
\\
\sin^2 2 \theta_{sol-SMA} &=& \{(0.001 - 0.004), 0.0015\}  \nonumber
\\
\sin^2 2 \theta_{LSND} &=& (0.001 - 0.04) . \nonumber
\end{eqnarray}
The LSND result will be used only in the four generation scenario with
a sterile neutrino. In the three generation case there are three
mixing angles. One of them is determined by the atmospheric
neutrinos, the second angle is determined by the solar neutrinos. For
the third angle the CHOOZ constraint is considered:
$$ \Delta m^2_{eX} < 10^{-3} ~ {\rm eV}^2 ~~~~~ {\rm or} ~~~~~
\sin^2 2 \theta_{eX} < 0.2  $$
We shall not consider any effect of CP violation and hence the mass
matrices are assumed to be real.

For a systematic analysis, we start with the possible textures for
the mass matrices presented in ref. \cite{alt}.
We consider all the texture mass matrices, which can explain the
maximal mixing for the atmospheric neutrinos and add possible
perturbations. Similar to the maximal mixing in the atmospheric
neutrino solution, the neutrinoless double beta decay imposes the next
strongest constraint on the general structure of the mass
matrices. Since the contribution to the neutrinoless double beta decay
is larger or equal to the mass squared difference required by the
atmospheric neutrino anomaly, the $M^\nu_{11}$ must be of the order of
any other large elements in the mass matrix. This rules out all the
textures with the $(11)$ element 0. In a more general analysis it is
possible to consider a scenario with vanishing $(11)$ element for the
largest entry in the mass matrix, but for
the present purpose of obtaining a simple texture mass matrix
we shall not discuss this possibility.

This would mean that the hierarchical neutrino mass schemes are all
ruled out \cite{imp}. The large angle solutions to the solar neutrinos are also
not allowed when the first two mass eigenvalues have opposite
signs. As mentioned above, in a more general case it may be possible
to have this texture and still satisfy the LMA solution, but it is not
possible to have any simple form of the mass matrix.
This argument leaves us with three mass textures
$$
M^{A1}_\nu = m_0 \pmatrix{ 1&0 & 0 \cr 0 & 1/2 & 1/2 \cr 0 & 1/2 & 1/2 }
~~~~~~~~
M^{A2}_\nu = m_0 \pmatrix{ 1& 0 & 0 \cr 0 & 1 & 0 \cr 0 & 0 & 1}
$$
\begin{equation}
M^{A3}_\nu = m_0 \pmatrix{ 1& 0 & 0 \cr 0 & 0 & 1 \cr 0 & 1 & 0},
\end{equation}
which can allow for both LMA as well as the SMA solutions.
There are three other textures, which can only allow for SMA solutions,
they are
$$
M^{B1}_\nu = m_0 \pmatrix{ 1&0 & 0 \cr 0 & -1/2 & -1/2 \cr 0 & -1/2 & -1/2 }
~~~~~~~~
M^{B2}_\nu = m_0 \pmatrix{ -1& 0 & 0 \cr 0 & 1 & 0 \cr 0 & 0 & 1}
$$
\begin{equation}
M^{B3}_\nu = m_0 \pmatrix{ 1& 0 & 0 \cr 0 & 0 & -1 \cr 0 & -1 & 0}.
\end{equation}
$m_0$ is the overall scale in these mass matrices, which is determined
by the value of the neutrinoless double beta decay.
The $B$ solutions are not very interesting, since they cannot
give us LMA solutions.

Let us first study the texture $A1$. The simplest perturbation that
may be considered is
\begin{equation}
m_1^{A1} = m_0 \pmatrix{a & b_1 & b_2 \cr b_1 & 0 & 0 \cr b_2 & 0 & 0} .
\end{equation}
In this case it is possible to have solutions with $b_1=0$ and $b_2 =
b$ or $b_1=b$ and $b_2 = 0$, which are exactly equivalent. Different
choices for the parameters $a$ and $b$ can now give the SMA or the LMA
solutions. This three parameter mass matrix with non zero element
depends on the condition, $[m_1^{A1}]_{22} = [m_1^{A1}]_{33} =
[m_1^{A1}]_{23} = [m_1^{A1}]_{32} $. A few representative sets of the
parameters which allow the LMA and SMA solutions are:
\begin{eqnarray}
(1) SMA &:& m_0 = 0.05; ~~~ a= 0.001; ~~~ b= 0.00002; \nonumber \\
(2) LMA &:& m_0 = 0.05; ~~~ a= 0.003; ~~~ b= 0.002; \nonumber \\
(3) LMA &:& m_0 = 0.05; ~~~ a= 0.003; ~~~ b_1=b_2= 0.001; \nonumber
\end{eqnarray}
We discuss these solutions briefly. These solutions correspond to
the partial degenerate case, where two of the masses are degenerate.
In this case it is not possible to have a neutrinoless double beta
decay contribution more than the mass required for the atmospheric
neutrinos. So, we get an effective mass contributing to the
neutrinoless double beta decay to be $0.05$. In all the cases the
mass squared difference between the states containing the $\nu_\mu$
and $\nu_\tau$ is around $\Delta m^2_{atm} \sim 0.0025$ eV,
which is required for the
atmospheric neutrinos. The mixing angle for the atmospheric neutrinos
is maximal or almost maximal ($\sin^2 2 \theta_{atm} > 0.95$). In (1), the
mass squared difference for solar neutrinos comes out to be
$\Delta m^2_{sol} \sim 5 \times 10^{-6}$ eV$^2$ with a mixing angle
$\sin^2 2 \theta_{sol} \sim 0.001$. In (2), we get
$\Delta m^2_{sol} \sim 2.5 \times 10^{-5}$ eV$^2$ with a mixing angle
$\sin^2 2 \theta_{sol} \sim 0.33$. In (3), there are no zero
elements, but since $b_1=b_2$, we get exactly maximal mixing for
atmospheric neutrinos $\sin^2 2 \theta_{atm} = 1$. For the solar
neutrinos the solution is similar to the case (2),
$\Delta m^2_{sol} \sim 2.5 \times 10^{-5}$ eV$^2$ with a mixing angle
$\sin^2 2 \theta_{sol} \sim 0.33$.

For the texture $A2$, it is not possible to have any elements to be
vanishing. The atmospheric neutrino maximal mixing requires that the
$(23)$ and the $(32)$ elements are non-vanishing. For maintaining the
maximal mixing for atmospheric neutrinos, we also need the $(12)$ and the
$(13)$ elements to be equal.
Thus one of the possible perturbations to the mass matrix $A2$ is
\begin{equation}
m_1^{A2} = m_0 \pmatrix{a & b & b \cr b & 0 & b \cr b & b & 0} .
\end{equation}
The SMA solution is not possible with any simple form of the perturbations.
A representative set for the LMA solution is: $m_0=0.4$ eV, $
a=0.0003$ and $b=0.003$, which gives
$\Delta m^2_{sol} \sim 6.3 \times 10^{-5}$ eV$^2$ with a mixing angle
$\sin^2 2 \theta_{sol} \sim 0.91$ and  $\Delta m^2_{atm} \sim 0.0028$
eV with exactly maximal mixing.

The most interesting case comes out to be the $A3$ case. All the
results may be accomodated with one parameter perturbation
\begin{equation}
m_1^{A3} = m_0 \pmatrix{a & 0 & a \cr 0 & 0 & 0 \cr a & 0 & - 2 a} .
\end{equation}
Now the complete mass matrix $m^\nu=M^{A3}_\nu + m_1^{A3}$ will also have
a few zero elements ($m^\nu_{22}=m^\nu_{12}=m^\nu_{21}=0$). Although
the allowed range is not very wide, there exist solutions, like
$a=0.003$ with $m_0 = 0.4$ eV. The predictions are,
$\Delta m^2_{sol} \sim 2 \times 10^{-4}$ eV$^2$ with a mixing angle
$\sin^2 2 \theta_{sol} \sim 0.33$ and  $\Delta m^2_{atm} \sim 0.0023$
eV with almost maximal mixing. Similar results come from two
other possible perturbations with this $A3$ texture
\begin{equation}
m_0 \pmatrix{a & a & a \cr a & 0 & 0 \cr a & 0 & - 2 a}
~~~~~ {\rm and } ~~~~~ m_0 \pmatrix{a & 0 & a \cr 0 & -a & 0 \cr a & 0 & - a}
\end{equation}
With such simple structures it is not possible to have a wider range
of parameters or accomodate the SMA solution.

For the mass matrix with no vanishing elements, it is now possible to
give a simple parametrization which guarantees a maximal mixing for
the atmospheric neutrinos and gives the neutrinoless double beta decay
\begin{equation}
{\cal M}_\nu = m_0 \pmatrix{m_{ee} & a & a \cr a & b+c & b-c \cr a & b-c & b+c} .
\end{equation}
The neutrinoless double beta decay determines the element
$m_{ee}$. The textures $A1$, $A2$ and $A3$ are the limiting cases with
$a,c \ll b$; $a \ll b=c$ and $ a \ll b = -c$, respectively. The CHOOZ
constraint is satisfied and the solar neutrino solutions are
obtained with suitable choice of $a$ and $c$.

In the case of four generations the analysis becomes more
involved. Only the mass matrices with minimum number of parameters
have been studied in this context \cite{dp,4gen}.
The simplest of these models require four parameters \cite{dp},
while other four-generation mass matrices require five or more parameters
with two identical diagonal elements \cite{4gen}.
We generalise the simplest of these mass matrices \cite{dp}
to include the neutrinoless double beta decay result and in
addition get the correct mixing angle for the solar neutrinos
\cite{dp}. In the minimal version of this mass matrix only maximal
mixing was possible for the LMA solution. We present a mass matrix
with only 5 parameters, which can explain all the experiments
including the neutrinoless double beta decay and the LSND result.

The mass matrix can be written in the basis
$[\nu_e, \nu_\mu, \nu_\tau, \nu_s]$ as
\begin{equation}
{\cal M}_{4 \nu} = \pmatrix{m & 0& a & d \cr 0& c & b & 0 \cr
a&b&0&0 \cr d&0&0& -m} .
\end{equation}
We can further economise by identifying two parameters $m=d$,
making it effectively a four-parameter mass matrix.

We discuss the solution briefly. The parameter $m$ determines the
amount of neutrinoless double beta decay. The oscillation between the
states $\nu_e$ and $\nu_s$ explains the solar neutrino problem. The
mixing angle now becomes, $\sin^2 2 \theta_{sol} = d^2/(m^2+d^2)$.
A simple choice of $d=m$ gives $\sin^2 2 \theta_{sol} = 0.5$, which
is consistent with present data. Restricting ourselves to $c \ll b$
ensures a maximal mixing between $\nu_\mu$ and $\nu_\tau$, as required
by the atmospheric neutrinos. The mass
squared difference for the atmospheric neutrinos is given by $2bc$
and that for the LSND is $b^2 -d^2 -m^2$. There are no simple
expressions for the mass squared difference required for the solar
neutrinos and the mixing angle for the LSND result. Numerically these
predictions come out as required.
For completeness we present a representative set of
the values of these parameters (all parameters are in eV),
$  a=0.03; ~~~~ b = 0.6; ~~~~ c = 0.003; ~~~~ d=m=0.1; $
which gives,
$ m_{ee} = 0.1$ eV, \\
$\Delta m_{sol}^2 = 7 \times 10^{-5}$ eV$^2$,
$~~~\sin^2 2 \theta_{sol} = 0.5$, \\
$\Delta m_{atm}^2 = 0.0027$ eV$^2$,
$~~~\sin^2 2 \theta_{atm} = 1$, \\
$\Delta m_{LSND}^2 = 0.34$ eV$^2$,
$~~~\sin^2 2 \theta_{LSND} = 0.003$.\\ This appears to be the simplest
four-generation mass matrix with texture zeroes,
which can explain all experiments in neutrino physics.

We have not mentioned about the LOW, Just-so and Vacuum oscillation
solutions of the solar neutrino problem. It has been shown in ref
\cite{impn} that in a general analysis these solutions are ruled
out when the radiative corrections are considered in conjunction with
the present result on the neutrinoless double beta decay. However, in
a more detailed analysis \cite{vis} it has been pointed out that the
LOW solution is allowed in some special cases. But from the point of
constructing a simple form of the texture mass matrix we could not
find any solution, which allows these solutions. In the four
generation case the mass matrix we presented can allow for a vacuum
solution for some choice of parameters, similar to the earlier
analysis of ref \cite{dp}.

In summary, we try to obtain the simplest forms of the mass matrices
consistent with all the present experiments including the neutrinoless
double beta decay. We present possible mass matrices with vanishing
elements. We also present a simple form of the mass matrix
with four parameters, which can reproduce all possible
allowed mass matrices. We further present a four generation mass matrix
with five parameters, which can explain all the experiments.

\vskip .2in
{\bf Acknowledgements:} One of us (US) would like to thank the
Max-Planck-Institute f\"ur Kernphysik, Heielberg for hospitality.

\newpage
\bibliographystyle{unsrt}

\end{document}